\documentclass[%
 reprint,
 amsmath,amssymb,
 aps,
]{revtex4-2}

\usepackage{graphicx}
\usepackage{dcolumn}
\usepackage{bm}
\usepackage{natbib}
\usepackage{color}

\begin{document}

\preprint{APS/123-QED}
\title{The Effect of Super-spreader Events in Epidemics}

\author{Harisankar Ramaswamy}
 \email{hramaswa@usc.edu}
\author{Assad A Oberai}%
 \email{aoberai@usc.edu}

\author{Mitul Luhar}
 \email{luhar@usc.edu}
 
\author{Yannis C Yortsos}%
 \email{yortsos@usc.edu}
 
\affiliation{%
 Viterbi School of Engineering\\
 University of Southern California, Los Angeles, CA 90089.
}%




\date{\today}

\begin{abstract}
The spread of infectious epidemics is often accelerated by super-spreader events. Understanding their effect is important, particularly in the context of standard epidemiological models, which require estimates for parameters such as $R_0$. In this letter, we show that the effective value of $R_0$ in super-spreader situations is significantly large, of the order of hundreds, suggesting a delta-function-like behavior during the event. Use of a well-mixed room model supports these findings. They elucidate infection kinetic modeling in enclosed environments, which differ from the standard SIR model, and provide expressions for $R_0$ in terms of physical and operational parameters. The overall impact of super-spreader events can be significant, depending on the state of the epidemic and how the infections generated by the event subsequently spread in the community.
\end{abstract}

\maketitle

\section{Introduction}\label{sec:intro}
The spread of infectious epidemics, such as the one the world is experiencing today with COVID-19, is often associated with super-spreader events. These refer to congregations of people in high areal (spatial) density environments, largely without protective face covering, with high degree of proximity and high intensity in vocal interaction, but lasting over relatively short periods of time compared to the characteristic time of the spread of epidemics.
Significant work has been done to model the role of super-spreader individuals who are responsible spreading the disease to a disproportionately large number of individuals, either through enhanced viral shedding or social interaction \cite{lloyd2005superspreading,james2007event}. However,  additional work is needed to model super-spreader events, especially through the lens of the types of models that are used to model the spread of pandemics.
These have received considerable public attention recently, and are crucially relevant to high-density, high-activity, or longer-duration interactions in enclosed environments \cite{nishiura2020closed,chang2021mobility}, e.g. the operation of schools and dining or entertainment establishments. Understanding their impact and incorporating it in conventional models is important, including when the susceptible fraction of the population has decreased due to vaccinations. It is the objective of this letter to address this important subject. 

We focus on how super-spreader effects can be captured in conventional models, including determining the value of effective infection parameters, such as the well-known reproduction number $R_0$, during the event. We base our analysis on a recently developed model for the spread of epidemics \cite{ramaswamy2021comprehensive}, which is based on an analogy between epidemics and chemical reaction processes.  Consistent with the standard SIR framework, the model contains the basic three populations (species), namely, susceptible ($S$), infected ($I$) and recovered (and which also includes perished) ($R$). The formalism allows one to include vaccinated or other sub-populations. Although not central to the present study, we will make some related comments. 

\section{Methods and Results} 

The variables of interest are absolute and relative population areal densities, namely $\rho$ (number of people/area) and the population (species) fractions, $s$, $i$, and $r$ (defined as $\rho_n/\rho$ where $\rho_n$ is the density of species $n=s,i,r$, respectively). The corresponding conservation equations were derived in \cite{ramaswamy2021comprehensive}, and are expressed in dimensionless notation as follows:
\begin{eqnarray}
\frac{\partial s}{\partial t} + Da \nabla \cdot (\bm{v} s)  - C \nabla (\ln \rho ) \cdot \nabla s &=& \nabla \cdot ( C \nabla s) - R_0 (\rho, r) s i  \label{eq:s1} \\
\frac{\partial i}{\partial t} + Da \nabla \cdot (\bm{v} i)  - C \nabla (\ln \rho ) \cdot \nabla i &=& \nabla \cdot ( C \nabla s) + R_0 (\rho, r) s i   - i \label{eq:i1} \\
\frac{\partial r}{\partial t} + Da \nabla \cdot (\bm{v} r)  - C \nabla (\ln \rho ) \cdot \nabla r &=& \nabla \cdot ( C \nabla r) + i \label{eq:r1} \\
\frac{\partial \rho}{\partial t} + Da \nabla \cdot (\bm{v} \rho)  &=&   0 \label{eq:density_nondim}
\end{eqnarray}
where $s+i+r=1$. In the above, space is normalized by a characteristic length $L$ and time is normalized by a characteristic time $\Lambda^{-1}$, which is a measure of the kinetics of the recovery, assumed to equal 14 days for COVID-19. We have also defined a dimensionless Damkohler number $Da=\frac{U}{L \Lambda}$, a dimensionless diffusion number $C= \frac{D}{\Lambda L^2}$, and the rescaled velocity $\bm{v}$ based on a characteristic dimensional velocity $U$. 
We note in (\ref{eq:s1})-(\ref{eq:density_nondim}) the formulation in terms of areal densities, which is the appropriate formalism, rather than the number of individuals (as is done in standard SIR-type models). 

Equations (\ref{eq:s1})-(\ref{eq:density_nondim}) include spatial transport by advection (through $\bm{v}$) and diffusion (through $C$), and reaction (conversion of one species to another). We model infection and recovery in terms of two irreversible chemical reactions between the species
\begin{eqnarray}
S + I &\mapsto& 2 I \label{eq:reac1} \\
I &\mapsto& R \label{eq:reac2}
\end{eqnarray}
The stoichiometric coefficient 2 on the RHS of (\ref{eq:reac1}) states that a new infected species $I$ is produced as a result of an interaction with susceptible species $S$, the net result being the creation of a new memmber of $I$. The latter is converted to the recovered (or perished) species $R$, in an one-to-one stoichiometry (\ref{eq:reac2}). Notably, the respective reaction rates in (\ref{eq:s1})-(\ref{eq:r1}) follow typical mass action kinetics.

The infection intensity is characterized by the important dimensionless parameter  
\begin{eqnarray}
R_0(\rho,r) = \frac{K_0 \rho }{\Lambda} \kappa(\rho, r) \label{eq:defrho1} 
\end{eqnarray}
which is proportional to the density $\rho$, and to the kinetic parameters $K_0$ (dimensions of [time$\times$(number/area)]$^{-1}$) and $\kappa(\rho, r)>0$ (dimensionless). 
Parameter $K_0$ implicitly accounts for the frequency of encounters (collisions) between individuals and the density of interaction, in addition to biological (infection) and environmental (e.g., face covering) factors. The dimensionless parameter $\kappa(\rho, r)$ varies with spatial density as well as the extent of the epidemic, $r$. It is important for modeling the spreading of infection in open environments as it captures effects of spatial distancing. For example, beyond a critical distance (below a critical density value $\rho_0$) infection rates are negligible ($\kappa(\rho, r) \ll 1$); conversely, there is a maximum limit $\rho_1$ denoting closest packing. 

Equations (\ref{eq:s1})-(\ref{eq:density_nondim}) were solved in \cite{ramaswamy2021comprehensive} for a variety of conditions, from the case of a ``batch reactor'', when transport is fast and all spatial gradients vanish (and where the resulting equations lead to the conventional SIR model), to cases that include mobility (via diffusion and/or advection of the population species). The latter case leads to the propagation of spatial waves, either through diffusion or through macro-advection (which is similar to turbulent dispersion). 

It is important to anticipate that while the rate expressions in (\ref{eq:s1})-(\ref{eq:r1}) depend linearly on $i$ in the product $is$, this might  not necessarily be representative of the kinetic rates in enclosed rooms, where air circulation and the biology of infection cause longer-range interactions, potentially giving rise to behavior not captured by (\ref{eq:reac1}). This is an important point, further discussed in Section~\ref{sec:R0_SS}. One signal of such a different behavior is the possibility that the $R_0$ of the conventional SIR model  does not remain constant (i.e., is time-dependent) during indoor super-spreader events.

\subsection{SIR Model for Super-spreader Events}\label{sec:SIR_SS}

Assume a ``batch reactor'' model (zero spatial gradients), in which case the governing equations reduce to the equivalent of a more traditional SIR model \cite{kermack1927contribution,anderson1979population},  
\begin{eqnarray}
\dot{s} (t) &=& -R_0 (\rho,r) s i \label{eq:s2} \\
\dot{i} (t) &=&  R_0 (\rho,r) s i - i \label{eq:i2} \\
\dot{r} (t) &=&  i \label{eq:r2}
\end{eqnarray}
subject to  
\begin{eqnarray}
s + i + r &=& 1  \label{eq:clos1} 
\end{eqnarray}
and the initial conditions
\begin{eqnarray}
i(0)=i_0, \qquad s(0) \equiv s_0=1-i_0, \qquad  r(0)= 0. \label{eq:ic1} 
\end{eqnarray}
The problem involves two dimensionless parameters, $i_0$ and $R_0$. The latter parameter is the reproduction number, typically interpreted as the average number of people an infected person will infect. Significantly, this interpretation does not clarify the interval of time over which such infection will occur. We can make a simple calculation to provide estimates. By ignoring the dependence of $R_0$ on $r$, and taking $s$ constant, (\ref{eq:s2})-(\ref{eq:ic1}) give  
\begin{eqnarray}
i \approx i_0 \exp \left[(R_0-1)t\right] 
\label{eq:exp} 
\end{eqnarray}            
which is the expected exponential rise in infections at the onset of the epidemic if $R_0>1$. Then, the time interval, $t_{R_0}$, required for $R_0$ to be interpreted as the average number of new infections caused by an average infected person, $R_0 = i/i_0$, is 
\begin{eqnarray}
t_{R_0}=\frac{\ln(R_0)}{R_0-1}
\label{super}
\end{eqnarray}
For example, for $R_0=2$, $4$ or $250$ we find $t_2=0.69$,  $t_4=0.46$ or $t_{250}=0.022$ namely, about 10 days, 7 days, or 7.45 hours, respectively, with the present normalization for time. Of course, more refined estimates that capture the effect of the decrease of $s$ and $R_0$, etc., are possible, given that the infection curve ceases to be exponential after some time. 

Qualitative as it might be, this observation is significant in relation to super-spreader events. Typically, these occur over a time of the order of a few hours, which in the present dimensionless notation, expressed with a characteristic time of about two weeks, is very small (e.g., roughly 0.015 for an event of 5 hrs duration). It follows that for any significant infection to occur during these events, the corresponding value of $R_0$ must be large during that time interval. Although consistent with the super-spreader terminology, the finding that $R_0$ must be significantly large is novel and is  explored further in this letter.  

Assume now, that with the epidemic following equations (\ref{eq:s2})-(\ref{eq:ic1}), a super-spreader event occurs at time $t^*$, and that it lasts over a small interval of dimensionless time $\epsilon$. The solution of the SIR problem up to time $t^*$ \cite{ramaswamy2021comprehensive} provides the initial conditions to the super-spreader event, namely
\begin{eqnarray}
t^* = \int_0^{r^*} \frac{du}{1 - u - s_0 \exp(- \int_0^u R_0(r') dr') }
\end{eqnarray}
from which we have 
\begin{eqnarray}
s_{-} &=& s_{0} \exp \left[ - \int_0^{r^*} R_0(r) dr \right]    \nonumber \\  i_{-} &=& 1 - s_{-} - r^{*} ; c^{*}\equiv 1-r^{*}= s_{-}+ i_{-}   
\label{abc}
\end{eqnarray}
where superscript $*$ denotes the event, $r^*=r(t^*)$ is the recovered fraction, $c^{*}$ is the total concentration, and subscript $-$ denotes conditions just prior to the event. Next, we assume that the SIR model applies during the event, and rescale time around $t^*$ by defining $\tau=(t-t^*)/\epsilon$, where $\epsilon \ll 1 $ is a measure of the duration of the event. Allowing $R_0$ to depend on time, equations (\ref{eq:s2})-(\ref{eq:ic1}) become 
\begin{eqnarray}
\frac{ds(\tau)}{d \tau}  &=& -\epsilon R_0 (t^* + \epsilon \tau , r^* ) s i \label{eq:s3} \\
\frac{di(\tau)}{d \tau}  &=&  \epsilon R_0 (t^* + \epsilon \tau , r^* ) s i - \epsilon i \label{eq:i3} \\
 \frac{dr(\tau)}{d \tau}  &=&  \epsilon i \label{eq:r3}
\end{eqnarray}
In the small $\epsilon$ limit, these have a non-trivial solution only if the product $\epsilon R_0 (t^* + \epsilon \tau , r^* )$ is finite, thus suggesting indeed a large $R_0$ (and a delta-function-like behavior) during the event. As expected, (\ref{eq:r3}) gives a constant $r$ (there are no new recoveries during the small-duration event), hence 
\begin{eqnarray}
s_{+}+i_{+}= s_{-}+i_{-}=c^*
\end{eqnarray}
Here, subscript $+$ denotes the state after the conclusion of the event, and the sum $c^*$ of susceptible and infected individuals is constant just before, during and immediately after the event. Equation~(\ref{eq:i3}) now reads 
\begin{eqnarray}
\frac{d i (\tau)}{d\tau}=\epsilon R_0 (t^*+\epsilon \tau;r^* )(c^*-i)i
\label{fund}
\end{eqnarray}
which can be integrated to yield
\begin{eqnarray}
\frac{is_-}{i_ - (c^*-i)} = \exp \left[ c^* \epsilon \int_0^\tau R_0 (t^*+ \epsilon \tau;r^* ) d\tau \right]
\end{eqnarray}
If the conclusion of the super-spreader event is at $\tau=\tau_\infty$, we can define the integral
\begin{eqnarray}
\epsilon \int_0^{\tau_\infty} R_0 (t^*+ \epsilon \tau;r^* ) d\tau \equiv b
\end{eqnarray}
which is the cumulative action of $R_0$ during the event. This “infection action” parameter will be expressed in terms of measurable parameters.   

It follows that the standard SIR type model during the super-spreader event leads to 
\begin{eqnarray}
i_+ = \frac{i_-  c^*  \exp[c^* b]}{s_- + i_-  \exp[c^* b]}
\label{eq:i+1}
\end{eqnarray}
for the infected fraction at the conclusion of the event and 
\begin{eqnarray}
s_+ = \frac{s_-  c^*}{s_- + i_-  \exp[c^* b]}\label{eq:s+1}.
\end{eqnarray}
for the susceptible fraction. As expected,  $s_+<s_-<1$ and $i_+>i_-$, with the maximum value of $i_+$ equal to $c^*$ when $s_+=0$ (all susceptible individuals are infected). If 
$c^*b \ll 1$, the event is not of the super-spreader kind, and $i_+=i_-$  and $s_+=s_-$. This is not so for a large super-spreader type event. Note that (\ref{eq:i+1}) and (\ref{eq:s+1}) are valid regardless of the evolution of the epidemic prior to the super-spreader event (they are independent of the use of (\ref{eq:s3})-(\ref{eq:r3})), as the only inputs prior to the event are $i_-$ and $s_-$.      

\begin{figure}
\centering 
\includegraphics[width=0.6\linewidth]{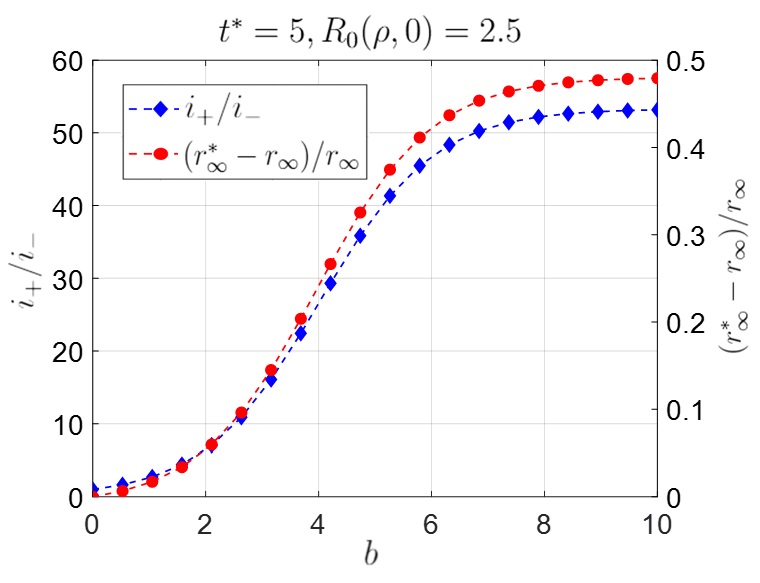}
\caption{\label{fig:varb} The ratio of infections at the end and the beginning of a super-spreader event, and the corresponding effect on the change in asymptotic herd immunity, as a function of the infection action parameter $b$ for a fixed value of $t^*=5$ and $R_0 (\rho,0)=2.5$ (and where $r_\infty=0.675$).}
\end{figure}

The effect of $b$ on the ratio $i_+/i_-$ is shown in Figure \ref{fig:varb}, for $t^*=5$, $r^*=0.0127$, $i_0=10^{-5}$ (relatively early stages of an epidemic) and $R_0 (\rho,0)=2.5$. As $b$ increases, the infection ratio initially increases almost exponentially, but asymptotically saturates to the limiting value
\begin{eqnarray}
\frac{i_+}{i_-} \approx 1 + \frac{s_-}{i_-},       
\end{eqnarray}
where $s_+ \approx 0$. 
The effect is larger near the beginning of the epidemic (small $t^*$) when $i_- \ll 1$ and the infection ratio is approximately exponential over a substantial interval
\begin{eqnarray}
\frac{i_+}{i_-} \approx \exp [c^* b]
\end{eqnarray}
Also plotted in Figure \ref{fig:varb} is the relative increase in the value of the ``herd immunity" after the super-spreader event, $(r_\infty^*-r_\infty)/r_\infty$, where $r_\infty$ is defined as the final value of $r$ after a wave of the epidemic subsides at the prevailing value of $R_0 (\rho,0)$ (see Section~\ref{sec:overall} as to how $r_\infty$ is calculated). This increase is calculated assuming that the newly infected populations from the event return to the populations from which they originated and at the same densities (see Section~\ref{sec:overall}). The increase in $r_\infty$ follows a similar pattern to the increase in the infection ratio, ultimately leading, for the conditions of Figure \ref{fig:varb}, to the theoretical maximum $r_\infty^* \approx 1$.

\begin{figure}
\centering 
\includegraphics[width=0.6\linewidth]{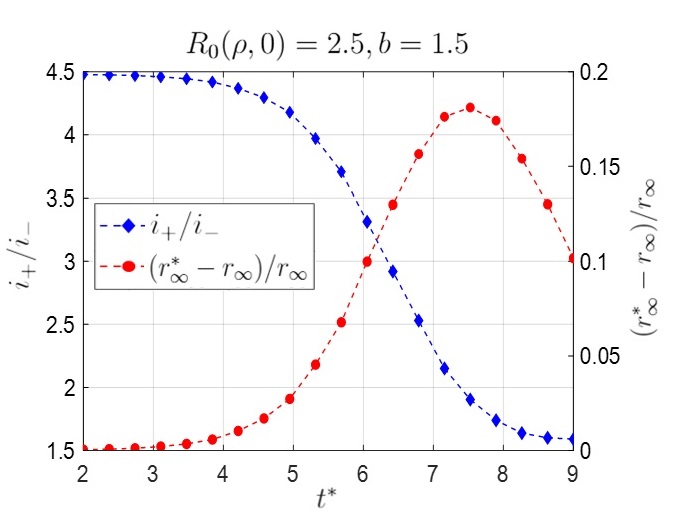}
\caption{\label{fig:vart} The ratio of infections at the end and the beginning of a super-spreader event, and the corresponding effect on the change in the asymptotic herd immunity as a function of the timing of the event for a fixed value of $b=1.5$ and $R_0 (\rho,0)=2.5$. }
\end{figure}

The effect of the timing of the event, $t^*$, is shown in Figure \ref{fig:vart}, for the relatively small value of $b=1.5$, $i_0=10^{-5}$, and $R_0 (\rho,0)=2.5$. 
At small values of $t^*$ the increase in infections due to the event is  equivalent to simply increasing the initial fraction of infected individuals. As discussed in \cite{ramaswamy2021comprehensive}, an increase in the initial number of infections acts to accelerate the onset of the epidemic, but without a significant effect on the value of $r_\infty$. (Of course, delaying the onset of the epidemic can help raise public awareness and enable the adoption of preventive measures that could lead to different behavior---a smaller $R_0 (\rho,0)$---and hence lower infections.) At larger values of $t^*$, when the epidemic has been evolving for a while, the super-spreader event acts as an additional and significant promoter of infection, resulting in non-trivial increases in the fraction of infected populations and the asymptotic ``herd immunity" at the given $R_0 (\rho,0)$. We note the existence of a maximum effect around $t^*=7.5$, where the fractional increase in $r_\infty$ reaches as high a value as 17\%. Overall, the super-spreader event acts as a catalyst that further enhances the contagion. 
\begin{figure}
\centering 
\includegraphics[width=0.6\linewidth]{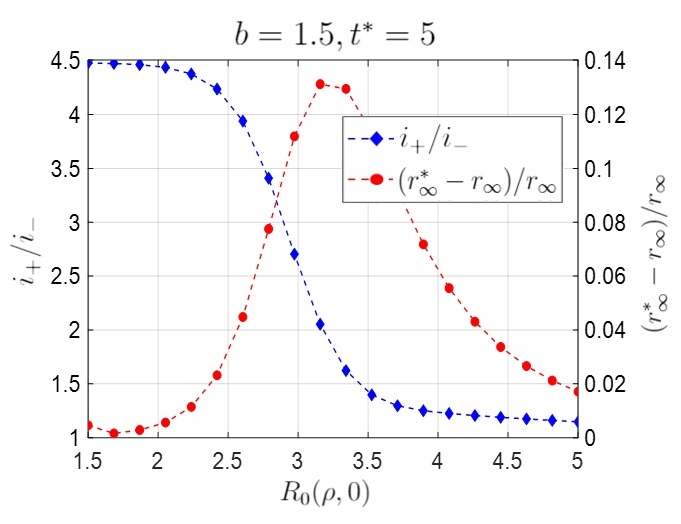}
\caption{\label{fig:varr} The ratio of infections at the end and the beginning of a super-spreader event, and the corresponding effect on the change in the asymptotic herd immunity as a function of $R_0 (\rho,0)$, for $t^*=5$ and $b=1.5$.}
\end{figure}
A similar behavior is observed for the effect of $R_0 (\rho,0)$ (Figure \ref{fig:varr} for $t^*=5$ and $b=1.5$). The relative increase in the asymptotic herd immunity is notable, increasing with $R_0 (\rho,0)$ and reaching a maximum of about 13\% at the intermediate value of $R_0 (\rho,0)=3.2$, following which it declines as $R_0 (\rho,0)$ further increases. This maximum, and the subsequent decline, suggest that when the (ambient) value of $R_0 (\rho,0)$ exceeds a certain threshold, which depends on $t^*$ and $b$, the intensity of infection is already large, and additional “flare-ups” have relatively negligible impact.

\subsection{$R_0$ During a Super-spreader Event}\label{sec:R0_SS}

We now turn to the question of what is the effective value of $R_0$ during a super-spreader event. We will follow \cite{luhar2020airborne}, who developed a quantitative model for airborne viral emission and COVID-19 transmission risk in enclosed rooms, using a well-mixed room model \cite{nazaroff2016indoor}. In the context of airborne disease transmission, such models are typically referred to as {\it Wells-Riley} models \cite{wells1934air,riley1978airborne,nicas2005toward,noakes2009mathematical,sze2010review}, and have recently been used to calculate COVID-19 transmission risks in specific case studies \cite{miller2020transmission,buonanno2020estimation}.   

For an event of duration $T$ (in dimensional notation), \cite{luhar2020airborne} provides the following expression for the increase in the infection fraction 
\begin{eqnarray}
i_+ - i_- = p_t s_{-}
\label{probab}
\end{eqnarray}
thus, 
\begin{eqnarray}
s_+ = c^* - i_+ = (1-p_t)s_{-}\label{eq:s+2}
\end{eqnarray}
where the “transition probability” $p_t$  is defined as
\begin{eqnarray}
p_t = 1- \exp\left(-\frac{D}{D_i}\right)
\label{kinet}
\end{eqnarray}
Here, $D$ denotes the cumulative viral intake, namely the dose, in units of number of RNA copies, of any individual at the event, over the time interval of the event and $D_i$ is the intake (dose) that leads to transmission in about 63\% of the cases. 

To a good approximation \cite{luhar2020airborne}, the dose can be estimated as follows
\begin{eqnarray}
D &\approx& \rho i_- (1-\hat{f})(1-f) \frac{Q}{\Lambda h E_{vac}} p_a QC_a^i \Delta t_\infty \label{eq:D}
\end{eqnarray}
where parameters $f$ and $\hat{f}$ represent the effectiveness of facial coverings, $Q$ is the volumetric respiration rate, $E_{vac}$ is the number of room air volume exchanges per time, $h$ is room height, $p_a$ is the fraction of time an initially infected person is active (i.e., speaking loudly), $C_a^i$ is the viral concentration emitted from the active infected person, and $\Delta t_\infty=\Lambda T$ is the dimensionless duration of the event, typically a small number. 
Implicit to the above are conditions of steady state, a settling time scale much larger than the air turnover time, and no information on the decay rate for the virus.

Before proceeding further it is worth discussing equations (\ref{probab}) and (\ref{kinet}). Rewrite (\ref{probab}) as 
\begin{eqnarray}
\frac{\Delta i}{\Delta t} =F(i,\Delta t)is
\label{propto}
\end{eqnarray}
where 
\begin{eqnarray}
F(i, \Delta t)\equiv \frac{p_t}{i\Delta t }
\label{frac}
\end{eqnarray}
Essentially, the RHS of (\ref{propto})  is the kinetic rate expression for the conversion of population $s$ to $i$. In the ``dilute" limit $\frac{D}{D_i} \ll 1$, we have $p_t \approx \frac{D}{D_i} \propto i \Delta t$, thus, $F(i, \Delta t)= \rm{const}$, similar to the model in equations (\ref{eq:s1})-(\ref{eq:density_nondim}). In the more general case, however, when $\frac{D}{D_i}=O(1)$, the rate dependence is different. Then, the product $F(i,\Delta t)i$ effectively results in the lowering of the exponent on $i$ in (\ref{propto}), which in theory will tend to zero in the limit of large $\frac{D}{D_i}$, where $p_t \approx 1$ and $i_+ - i_- \approx s_-$, thus invalidating the SIR model assumptions. The explanation is that in an enclosed environment with considerable mixing, one infected individual can infect multiple (and potentially all) susceptible persons, regardless of proximity, thereby invalidating the one-to-one equivalence in (\ref{eq:reac2}). 

We can account for this by assuming a different reaction scheme, perhaps of the form
\begin{eqnarray}
S + \frac{1}{m}I &\mapsto& I + \frac{1}{m}I \label{eq:reacn}
\end{eqnarray}
where the stoichiometric coefficient $\frac{1}{m}$ ($m>1$) indicates that one infected species can infect more than one susceptible species, each of which is then converted to the infected population. In such a case, and assuming that the law of mass action kinetics still applies, the resulting reaction rate dependence will then become
\begin{eqnarray}
\rm {rate} \propto i^{\frac{1}{m}}s 
\label{eq:reacn2} 
\end{eqnarray}
In fact, in the limit $m\gg 1$ one could capture the results in  (\ref{probab})-(\ref{kinet}). Indeed, let's assume that in (\ref{fund}), the reaction rate dependence  on species $I$ is only through the initial relative density $i_-$, namely take    

\begin{eqnarray}
\frac{d i (\tau)}{d\tau}=\epsilon R_{0m} (t^*+\epsilon \tau;r^* )(c^*-i)i_-
\label{fund2}
\end{eqnarray}
where we used $R_{0m}$ to denote the corresponding value of $R_0$ under these conditions. This can be integrated to yield at the end of the event
\begin{eqnarray}
i_+-i_-= (c^*-i_-&)&\left[1-\exp[-b_mi_-]\right]= 
s_-\left[1-\exp[-b_mi_-]\right]
\label{fund3}
\end{eqnarray}
Then, a direct comparison with (\ref{probab})-(\ref{kinet}) shows that
\begin{eqnarray}
b_m=\frac{1}{i_-}\frac{D}{D_i} 
\label{fund4}
\end{eqnarray}
hence 

\begin{eqnarray}
R_{0m} = \rho (1-\hat{f})(1-f) \frac{Q}{\Lambda h E_{vac}} p_a \frac{QC_a^i}{D_i}. \label{eq:R02}
\end{eqnarray}
For a quantitative estimate for $R_{0m}$, we assume the following parameters consistent with known COVID-19 super-spreader events \cite{hamner2020high,miller2020transmission,lu2020covid}: no mask usage ($f=\hat{f} =0$), limited spatial distancing ($\rho \approx 0.25$ m$^{-2}$), average room height  ($h=2.5$ m), low ventilation rates ($E_{vac}=2$ hr$^{-1}$), high activity levels ($p_a=0.5$) with infectious persons emitting 100 quanta (infectious doses $D_i$) per hour (i.e., $\frac{Q C_i^a}{D_i} =100$ hr$^{-1}$, \cite{buonanno2020estimation,miller2020transmission}), moderate respiration rate ($Q = 0.8 \times 10^{-4}$  m$^3$s$^{-1}$), typical recovery kinetics ($\Lambda^{-1}=14$ days). Then, (\ref{eq:R02}) gives $R_{0m} \approx 242$. 

This is an important result, consistent with the expectation that super-spreader events lead to large values of the reproduction parameter, in this case $R_{0m}$. The relation obtained in (\ref{eq:R02}) is the first such relation to our knowledge. The almost delta-function-like behavior for $R_{0m}$ during the super-spreader event must be noted. Assuming further an event of duration 5 hrs, we find $b_m\approx 242\times0.015=3.63$, a value roughly midway in the range of Figure \ref{fig:varb}. Using the definition of $R_0$, we can also obtain an approximate expression for the effective kinetic parameter during the event
\begin{eqnarray}
K_0 \kappa \approx  (1-\hat{f})(1-f) \frac{Q}{\Lambda h E_{vac}} p_a \frac{QC_a^i}{D_i}. \label{eq:K0kappa}
\end{eqnarray}
Equation (\ref{eq:K0kappa}) relates the kinetics of the infection to physicochemical, physiological and event operational parameters.  

The above show that if one were to use the SIR model during a super-spreader event, it would need to be modified such that the rate dependence on the infected fraction is proportional to its initial value, rather than allowing it to depend on its current one. The reason for this rests on the fact that in enclosed environments of relatively small duration (order of hours), the infection spreads uniformly due to circulation, therefore it should only depend on the initial, rather than the current,  fraction. Indeed, it is also unlikely that a newly infected person during the event would be able to infect another one, in the short time of the event.

\begin{figure}
\centering 
\includegraphics[width=0.6\linewidth]{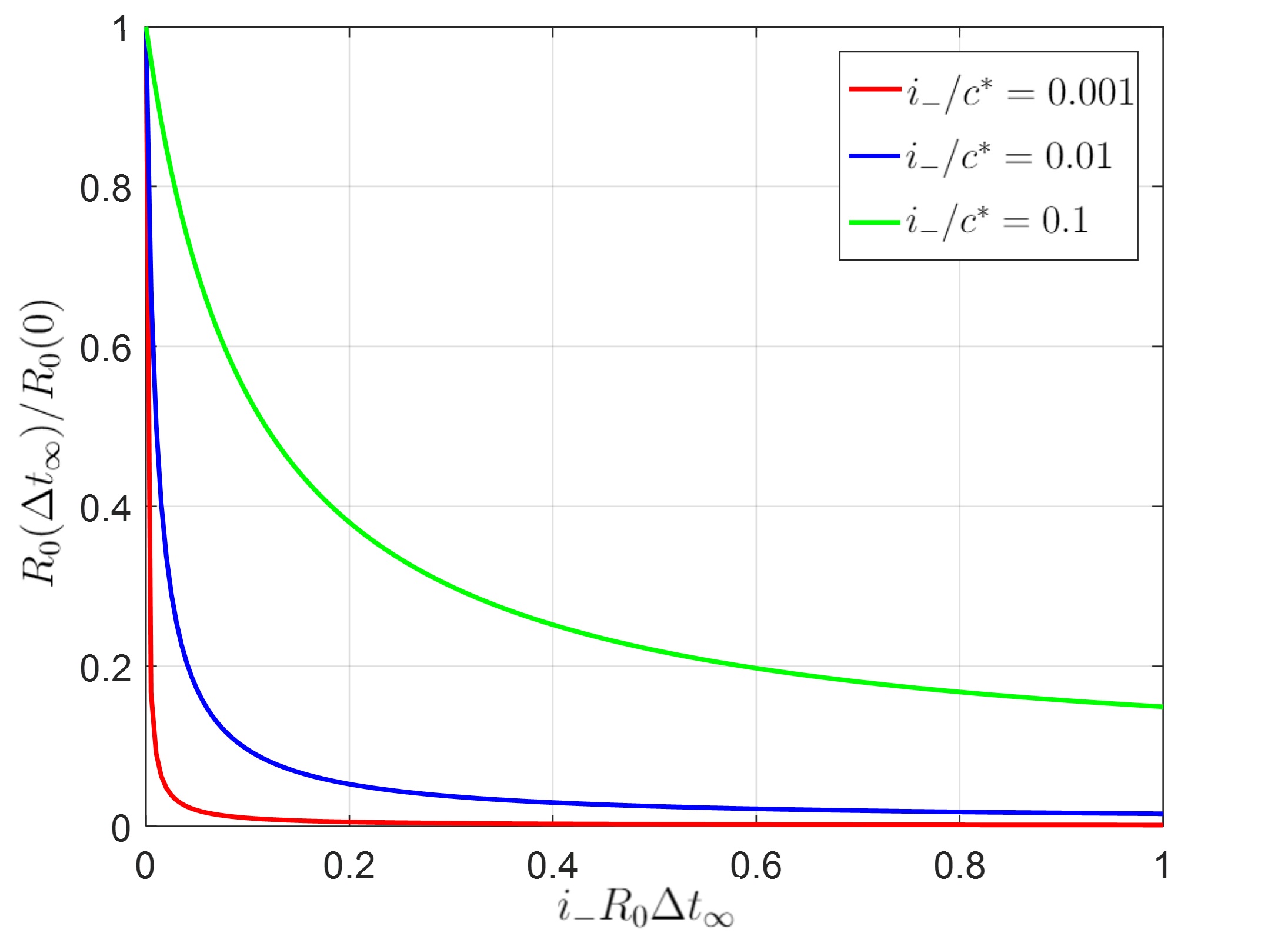}
\caption{\label{fig:vartinf} The effect of the duration of the event on the value of $R_0$, for three different values of $\frac{i_-}{c^*}$.} 
\end{figure}

Can the standard SIR model be used? Assuming that it is also valid during the event, we can combine (\ref{eq:s+1}), (\ref{eq:s+2}), and (\ref{kinet}) to find
\begin{eqnarray}
\exp[c^* b]=\frac{c^*}{i_-}\exp\left(\frac{D}{D_i}\right) - \frac{s_-}{i_-} \label{eq:c*b}
\end{eqnarray}
to yield the following expression for $R_0$  
\begin{eqnarray}
R_0= \frac{\frac{1}{D_i}\frac{D}{\Delta t_\infty}}{c^*-s_{-}\exp\left(-\frac{D}{D_i}\right)} \label{eq:R0}
\end{eqnarray}
At the onset of the event, we have $R_0(0)=R_{0m}$, in which case the SIR-like model captures correctly the infection action. Outside that limit, however, (\ref{eq:R0}) shows that $R_0$ will decrease with the duration of the event, which makes sense in that the SIR model kinetics depending on the combination $R_0 i$, rather than $R_{0m}i_-$, would require a smaller $R_0$ as $i$ increases. We can rewrite (\ref{eq:R0}) as follows  
\begin{eqnarray}
\frac{R_0}{R_0(0)} = \frac{\frac{i_-}{c^*}}{\left(1-\left(1-\frac{i_-}{c^*}\exp[-i_-R_0(0)\Delta t_{\infty}\right)\right)}
\label{eq:D_Di}
\end{eqnarray}
showing that $R_0$ is a decreasing function of the duration of the event (actually, of the product $i_-R_0(0)\Delta t_{\infty}$) and asymptotically stabilizes to the value

\begin{eqnarray}
\frac{R_0(\infty)}{R_0(0)} = \frac{i_-}{c^*}
\label{eq:D_D2}
\end{eqnarray}
The variation  of $R_0$ with time is shown in Figure \ref{fig:vartinf} 
for various values of the ratio $\frac{i_-}{C^*}$. The fact that $R_0$ is not a constant, but varies with time, is a demonstration that the infection kinetics during the event are not accurately captured with the linear dependence of the reaction rate on $i$, as assumed in the SIR-type model, but rather require the different interpretation along the lines suggested in (\ref{fund2}).  
       
One concludes that use of the standard SIR model under super-spreader events will lead in general to a value of $R_0$ that depends on the event duration suggesting that the enclosed-room kinetics may not be properly captured. The appropriate approach for such cases would be to keep the dependence on $I$ fixed at the initial infection conditions, in which case one obtains a revised and more appropriate constant value, $R_{0m}$.

\subsection{Overall Impact}\label{sec:overall}

The overall impact of the super-spreader event depends substantially on the conditions that follow it and specifically how the attendees of the super-spreader event disperse. Denote by $N_E$ the total number of attendees of the event, and by $N_T$ the total number of people from which the specific subset $N_E$ emanated and to which they return following the event. The evolution of infections is still described by  equations (\ref{eq:s2})-(\ref{eq:clos1}), integrated for $t>t^*$, but now subject to the new initial conditions at $t^*$, namely
\begin{equation}
i(t^*) = i_{+} \mu + i_{-} (1-\mu) ; s(t^*) = s_{+}  \mu + s_{-} (1-\mu) ; 
\end{equation}
where we introduced the ratio $\mu \equiv N_E/N_T$. The largest impact will be felt when $\mu \approx 1$, i.e., when the super-spreader event is attended by almost the entire community. Conversely, the effect will be least when $\mu \ll 1$. We explore these two limits next.

For $\mu \approx 1$, we have 
\begin{eqnarray}
t = \int_{r^*}^{r} \frac{du}{1 - u - s_+ \exp(- \int_{r^*}^u R_0(r') dr')} + t^*,
\end{eqnarray}
from which we can calculate 
\begin{eqnarray}
s &=& s_{+} \exp(-\int_0^{r^*} R_0(r') dr') \\
i &=& 1-r-s_{+} \exp(-\int_0^{r^*} R_0(r') dr')
\end{eqnarray}
A most interesting effect is on the final value of $r$ (which can be also defined as the herd immunity, $r_\infty$, corresponding to the final value of $R_0$). In the absence of a super-spreader event, this is the solution of
\begin{equation}
1-r_{\infty}-s_0  \exp(-\int_0^{r_\infty} R_0(r') dr') = 0.
\end{equation}
However, under conditions of a super-spreader event, and for $\mu \approx 1$, the corresponding asymptotic value becomes instead the solution of 
\begin{equation}
1-r_{\infty}^* = s_{+}  \exp(-\int_{r^*}^{r_\infty^*} R_0(r') dr')
\label{final}
\end{equation}
Under these conditions, the impact of the super-spreader event on the total number of new infections would be the difference 
$r_\infty^*-r_\infty$, corresponding to the normalized increase $(r_\infty^*-r_\infty)/r_\infty$. The effect of the various parameters on the new asymptotic value was considered in Figures \ref{fig:varb}-\ref{fig:varr}.

The effect is very different when $\mu \ll 1$. In this case, the infected individuals from the super-spreader event act as nucleation sites for new infections as they disperse back to their communities. This is particularly important when they join communities where infection was not originally present, and where, even though the corresponding initial infection fraction is small, it would be sufficient to initiate a new infection epidemic, although delayed by a time corresponding to the size of the initial fraction, as showed in \cite{ramaswamy2021comprehensive}. There, we also showed that while the onset of a new infection is delayed by $t_0 = -\frac{\log (i_{+} \mu ) \ln(10)}{R_0(0)-1}$ it occurs nonetheless, with a shape that is practically independent of the initial condition.   

We close with a note regarding the effect of additional super-spreader events, under the condition $\mu \approx 1$, where the previous analysis holds. Because of the continuous spread of the contagion, new super-spreader events will cause progressively smaller values of $s_+$, which in turn will lead to larger contagion, and ultimately to $r_\infty^*\approx 1$.


\section{Conclusions}\label{sec:conclusion}

In this manuscript we focused on obtaining an understanding of how super-spreader events can be quantitatively described. Using a recent model, applied to super-spreader conditions, we show that the impact of such an event is significant only if the conventional, widely used, parameter $R_0$ takes significantly large values, of the order of several hundred, during the event. In many ways, the variation of $R_0$ is delta-function-like during the event. We used a well-mixed room model and a different rate dependence on the infected fraction to demonstrate that such behavior is indeed observed, and to relate effective parameter values, as well as the corresponding kinetic parameters, to measurable physicochemical, physiological and operational parameters. Modifying the kinetic rate expression in enclosed environments is necessary, as in such cases, the assumption that one infected person infects one susceptible person does not hold, and neither does the implicit assumption that newly infected individuals can infect others during the short duration (order of hours) of the event. The kinetic model we used captures the ability for one infected person to infect multiple susceptible ones in enclosed environments. We then considered the effect of a super-spreader event on the spreading of the epidemic. While a super-spreader event acts to nucleate new infections in regions where such infections were originally absent, it also adds non-trivially to the spread of infections, even for cases where the epidemic has been on-going.

We close with a remark related to vaccinations. Assuming a significant vaccination extent, the overall effect in a super-spreader event will be related to two parameters, namely $c^*$ and $i_{-}$ both of which will be small. The value of $c^*$ affects the infection action parameter in equations (\ref{eq:i+1}) and (\ref{eq:s+1}). For example, if we take $c^*=0.2$ (namely $r^*=0.8$), the corresponding increase in the ratio of infections following the event will be equal to about 2.2 (where we have taken $b=4$) with minimal changes in the value of $s_+$, assuming a small value of $i_-$. Using (\ref{final}) we deduce that the overall impact will be very small. Additional work is needed to solidify these findings.

\section*{Contributions} YCY conceived the study and lead the efforts on the SIR aspects of the model. ML lead the effort on the well-mixed model. HR conducted the numerical simulations and generated the plots. AAO helped with mathematical and computational models. All authors contributed to writing and reviewing the manuscript.

\bibliography{apssamp}
\end{document}